\RequirePackage[2018-12-01]{latexrelease}
\documentclass{gGAF2e}
\begin{document}

\jvol{00} \jnum{00} \jyear{2012} 

\markboth{Tailleux et al.}{An energetically and thermodynamically
consistent Boussinesq model}

\articletype{ARTICLE}

\title{An energetically and thermodynamically
consistent Boussinesq model}

\author{
   R\'emi Tailleux${\dag}$$^{\ast}$,
   \thanks{$^\ast$Corresponding author. Email: r.g.j.tailleux@reading.ac.uk
   \vspace{6pt}} 
   Thomas Dubos${\ddag}$
   \vspace{6pt} and 
   Benedict J. Hatton${\ddag}$
   \\\vspace{6pt}  
   ${\dag}$Department of Meteorology, University of Reading, Reading, UK
   \\
   ${\ddag}$Laboratoire de M\'et\'eorologie Dynamique, \'Ecole polytechnique, Palaiseau, France
   \\
   \vspace{6pt}\received{received, accepted, etc.} 
   }
\maketitle

\begin{abstract}
The Boussinesq approximation is a cornerstone of geophysical fluid dynamics, yet its thermodynamic and energetic underpinnings have remained ambiguous. In standard formulations, the links with the fully compressible Navier--Stokes equations are obscured, internal energy is only implicit, and the representation of diffusion and irreversibility remains \textit{ad hoc}. Here we derive a new Boussinesq model in a fully traceable way from the two-component compressible Navier-Stokes equations, ensuring exact energy conservation and consistent thermodynamics. Assuming a linear equation of state, our model treats density as a proxy for specific volume, distinguishes in-situ and potential temperature explicitly, and incorporates diffusive fluxes that homogenise the correct thermodynamic potentials, ensuring consistent non-negative entropy production. The result clarifies the status of gravitational potential energy, resolves ambiguities surrounding salinity--entropy coupling, and retains the small terms carriers of key information about thermodynamics. Alongside this, we introduce an exact thermodynamically soundproof (TS) model whose weak diabatic divergence highlights the role of compressibility effects in stratified turbulence. Together, these models provide a transparent framework that reconciles classical approximations with compressible energetics, offering better-defined pathways for analysing stratified mixing, mixing efficiency, and the energy budgets of geophysical flows.
\end{abstract}


\section{Introduction}
\label{sec:introduction} 
The Boussinesq approximation remains one of the most widely employed approaches for simplifying the Navier--Stokes equations in geophysical and astrophysical fluid dynamics. By filtering out fast acoustic modes, the approximation provides a ``soundproof'' description of fluid motion while retaining essential buoyancy effects. For the past century it has been the workhorse for theoretical analyses and direct numerical simulations of stratified turbulence and convection \citep{lappaIncompressibleFlowsBoussinesq2022}. Its success rests on the assumption that density variations are small relative to a constant reference value, so that density may be taken as constant everywhere except in the buoyancy term. This leads to a divergence-free velocity field and a substantial reduction of mathematical complexity, enabling efficient modelling of low-Mach-number flows in diverse settings, from atmospheric and oceanic circulations to astrophysical phenomena.

Despite this, several aspects of the energetic and thermodynamic description of the Boussinesq approximation remain unclear and controversial. Because internal energy is only implicit and decoupled from mechanical energy, the link between between the approximate system and the fully compressible Navier--Stokes equations is neither obvious nor transparent. Seminal studies on the energetics of stratified fluids using the standard Boussinesq model \citep[e.g.][]{wintersAvailablePotentialEnergy1995} have had to rely on the idea of an implicit internal energy reservoir without being able to write down its expression. As a consequence, questions remain regarding the extent to which the Boussinesq equations truly capture the energetics of stratified mixing, available potential energy, and dissipative balances \citep{tailleuxEnergeticsStratifiedTurbulent2009}.

Originating from works by \citet{oberbeckUeberWarmeleitungFlussigkeiten1879} and \citet{boussinesqMiseEquationPhenomenes1902}, the derivation of the standard Boussinesq approximation has traditionally been done through \textit{ad hoc} scale analysis \citep[e.g.][]{spiegelBoussinesqApproximationCompressible1960,drazinHydrodynamicStability1981} or formal asymptotic analysis \citep[e.g.][]{mihaljanRigorousExpositionBoussinesq1962,zeytounianJosephBoussinesqHis2003} of the dynamical equations. The result is a simplification of the Navier--Stokes equations achieved by neglecting small terms such that the dynamical feedback causing acoustic wave propagation is suppressed. 
However, the information lost during this process leaves the underlying thermodynamic structure obscured, and can lead to confusion when interpreting energy budgets, entropy production, or the role of small but fundamental terms (such as non-conservation terms in the temperature equation or the salinity-dependent part of entropy). \citet{mihaljanRigorousExpositionBoussinesq1962} suggests that in order to find a closed energy cycle of a fluid subject to the Boussinesq approximation, one must incorporate first-order terms that are omitted from the leading-order dynamical equations. Alternatively, approximations may be made directly to an energy quantity, from which the dynamics may be deduced \citep{cotterVariationalFormulationsSoundproof2014, tortUsualApproximationsEquations2014}. This approach helps to ensure that terms key to a consistent energetic description are not lost.

In this paper, we take the latter approach based on approximation to the energetics, showing that many small and apparently negligible terms are nevertheless the carrier of important information about the fluid thermodynamics and energetics. While from the viewpoint of approximation theory, neglecting small terms may seem justified, the problem is that as the result of doing so, the resulting approximate system of equations contain less information than the non-approximated equations. A key question is whether this matters or not. For instance, the calculation of upper bounds on various quantities of interest in turbulent stratified mixing, such as horizontal convection \citep{paparellaHorizontalConvectionNonturbulent2002,wintersAvailablePotentialEnergy2009}, relies on the analysis of an energetically inconsistent Boussinesq model. Retaining more thermodynamic information in the Boussinesq approximation, and guaranteeing a closed energy budget may be the key to improving these bounds.

More general soundproof models such as the pseudo-incompressible and anelastic approximations have been shown to admit thermodynamically consistent formulations \citep{pauluisThermodynamicConsistencyAnelastic2008, kleinThermodynamicConsistencyPseudoincompressible2012}, and progress on the consistency of the Boussinesq approximation has previously been made by \citet{ingersollBoussinesqAnelasticApproximations2005, youngDynamicEnthalpyConservative2010}. Recently, a unified framework of thermodynamically and energetically consistent soundproof models has been developed \citep{eldredThermodynamicallyConsistentSemicompressible2021, tailleuxSimpleTransparentMethod2024}, including treatment of diabatic effects. In this framwork, the standard Boussinesq approximation is seen to be a special case of the more general anelastic approximation but with a constant reference density $\rho_0$ and a pressure-independent linear equation of state. \citet{tailleuxSimpleTransparentMethod2024} perform the approximations on the static energy $\Sigma$, and derive the dynamics through its derivatives, thus guaranteeing energetic consistency. In this way, the approximated models retain an explicit energy quantity that is traceable to that of the non-approximated Navier-Stokes equation.

The central aim of this work is therefore to construct a new Boussinesq model that is fully consistent with energetic and thermodynamic principles. Our approach is based on two steps. 
First, we formulate a complete thermodynamic description of a two-component stratified fluid with a linear equation of state, in which all relevant thermodynamic potentials are made explicit.
To this we apply an approximation to the static energy, following the framework of \citet{tailleuxSimpleTransparentMethod2024}, thus deriving an improved Boussinesq model that preserves key energetic relationships and clarifies the roles of in-situ temperature, potential temperature, and relative chemical potential. This new formulation not only ensures that irreversible diffusion of heat and salt is represented consistently, but also restores the transparent connection between the approximate model and its fully compressible counterpart. Key innovations include treating density as a proxy for specific volume, embedding pressure gradients in salt diffusion to homogenize relative chemical potential, and resolving long-standing ambiguities in gravitational potential energy (GPE). 

In parallel, we introduce another model where no approximation is made to the structure of the dynamical equations (retaining the standard compressible Navier-Stokes formulation), but which is endowed with the same simplified thermodynamics as the Boussinesq model (i.e. a pressure-independent linear equation of state). The lack of pressure dependence in the equation of state renders this model soundproof, and thus we shall refer to this as the thermodynamically soundproof (TS) model. The resulting dynamics include a weakly divergent velocity field which is diabatic in nature and grows with turbulence intensity, offering new insight into the energetics of stratified turbulence. Retaining some kind of compressible effect is often underappreciated in soundproof approximations, but is discussed in relation to the pseudo-incompressible approximation \citep{durranImprovingAnelasticApproximation1989, kleinThermodynamicConsistencyPseudoincompressible2012} of which the TS model is also a special case. The TS model occupies a middle ground between the fully compressible and Boussinesq systems, and helps to disentangle the effects of thermodynamic simplifications (i.e. linearising the equation of state) from those of ``soundproofing'' approximations, while highlighting how small terms carry fundamental information exploitable for deriving constraints on fluxes and mixing efficiencies.

The contributions of this paper can be summarised as follows:
\begin{enumerate}
    \item We clarify the thermodynamic structure underlying the standard Boussinesq approximation, identifying which terms are essential for maintaining irreversibility and energy conservation.
    \item We derive a new energetically and thermodynamically consistent form of the Boussinesq equations, in which density acts as a proxy for specific volume and diffusive processes are formulated on the correct thermodynamic variables.
    \item We construct the thermodynamically soundproof (TS) model, dynamically similar to the Boussinesq system but with a diagnostically divergent velocity, thereby exposing the explicit role of diabatic divergence in stratified mixing energetics.
\end{enumerate}

The paper is structured as follows. 
Section~\ref{sec:standard_boussinesq_model} reviews the structure and properties of the standard Boussinesq model, with the aim of highlighting the features to be improved. 
Section~\ref{sec:structure_nse} reviews the general principles underlying the construction of energetically and thermodynamically consistent soundproof approximations based on the static energy asymptotics framework of \citet{tailleuxSimpleTransparentMethod2024}. 
Section~\ref{sec:new_models} derives two new energetically consistent models based on simplified thermodynamics associated with a linear equation of state.
Section~\ref{sec:ie_gep_partition} justifies the partition of Boussinesq potential energy into gravitational and internal components using an energetic argument, with the TS model acting as a bridge between the exact and approximated equations.
Section~\ref{sec:conclusions} summarizes the contributions and discusses implications for geophysical fluid dynamics and beyond.

\section{Formulation and properties of the standard Boussinesq model}
\label{sec:standard_boussinesq_model} 

For clarity, we define the ``standard'' Boussinesq model for a two-component fluid to be the system
\begin{subequations}
\begin{gather}
   \frac{\mathup{D}{\bm v}}{\mathup{D} t} + 2 {\bm \varOmega} \times {\bm v} 
   + \frac{1}{\rho_{0}} {\bm \nabla} p = - \frac{\rho_b}{\rho_{0}} {\bm \nabla} \Phi 
   + \nu \nabla^2 {\bm v}, 
   \label{standard_boussinesq_momentum}
   \\
   {\bm \nabla} \,{\bm \cdot}\, {\bm v} = 0 ,
   \label{standard_boussinesq_continuity}
   \\
   \rho_b = \rho_{0} ( 1 - \alpha (\theta - \theta_0) + \beta (S - S_0) ),
   \label{standard_boussinesq_linear_eos}
   \\
   \rho_{0} \frac{\mathup{D}\theta}{\mathup{D} t} = - {\bm \nabla} \,{\bm \cdot}\, {\bm J}_{\theta}, 
   \qquad {\bm J}_{\theta} = -\rho_{0} \kappa_T {\bm \nabla} \theta , 
   \label{standard_boussinesq_tempearture}
   \\
   \rho_{0} \frac{\mathup{D} S}{\mathup{D} t} = - {\bm \nabla} \,{\bm \cdot}\, {\bm J}_S, 
   \qquad {\bm J}_S = -\rho_{0} \kappa_S {\bm \nabla} S ,
   \label{standard_boussinesq_salinity} 
\end{gather}
\end{subequations}
where ${\bm v}=(u,v,w)$ is the velocity field, $p$ is pressure, $\rho_{0}$ is the (constant) reference density, ${\bm \varOmega}$ is Earth’s rotation vector, $\Phi = g z$ is the geopotential, $\nu$ is kinematic viscosity. The density $\rho_b$ is a linear function of potential temperature $\theta$, and concentration $S$ of the secondary fluid component (hereafter referred to as ``salinity''). Constants $\theta_0$, $S_0$ are reference values of temperature and salinity, $\alpha$, $\beta$ are the thermal expansion and haline contraction coefficients, and $\kappa_T$, $\kappa_S$ are the molecular diffusivities of potential temperature and salt.

\subsection{Scope of the approximation} 
In its most general sense, the Boussinesq approximation applies only to the momentum and continuity equations (\ref{standard_boussinesq_momentum}--\ref{standard_boussinesq_continuity}), with the equation of state for $\rho_b$ and the diffusive fluxes \(\bm{J}_\theta\), \(\bm{J}_S\) left to be chosen. For example, numerical ocean models often employ realistic nonlinear equations of state including pressure dependence \citep{youngDynamicEnthalpyConservative2010,tailleuxSimpleTransparentMethod2024}, alongside parameterised heat and salt fluxes including skew and rotated diffusion components \citep{griffiesGentMcWilliamsSkew1998, griffiesIsoneutralDiffusionZCoordinate1998}. An alternative and often useful formulation of the momentum equation reads
\begin{equation}
   \frac{\mathup{D}{\bm v}}{\mathup{D} t} + 2 {\bm \varOmega} \times {\bm v} 
   + \frac{1}{\rho_{0}} {\bm \nabla} p' = b {\bm \nabla} \Phi 
   + \nu \nabla^2 {\bm v},
\end{equation}
where $p' = p - p_{\rm R}(\Phi)$ is the pressure anomaly relative to the hydrostatic Boussinesq reference pressure $p_{\rm R}(\Phi) = p_0-\rho_{0}\Phi$, and $b=-(\rho_b-\rho_{0})/\rho_{0}$ is the normalised buoyancy.

\subsection{Energetics} 
The energetics of the Boussinesq system is usually framed in terms of kinetic  and gravitational potential energies per unit mass,
\begin{equation}
   E_{\rm k} = \tfrac{1}{2}  \left\lvert {\bm v} \right\rvert ^2 , 
   \qquad 
   E_{\uphi} = \frac{\rho_b \Phi}{\rho_{0}}.
\end{equation}
The volume integral of $\rho_{0}E_{\uphi}$,
\begin{equation}
   \int_{V} \rho_{0} E_{\uphi} \, \ud V = 
   \int_{V} \rho_b \Phi \, \ud V ,
\end{equation}
formally coincides with the gravitational potential energy of a compressible fluid. Yet the analogy is subtle: in a compressible fluid the parcel mass is $\ud m = \rho_b \ud V$, whereas in a strictly Boussinesq fluid it is $\ud m = \rho_{0}\ud V$. This distinction complicates the interpretation of $E_{\uphi}$ and motivates closer scrutiny.

The evolution equations are obtained in the standard way: taking the dot product of (\ref{standard_boussinesq_momentum}) with ${\bm v}$ gives 
\begin{equation}
   \rho_{0} \frac{\mathup{D} E_{\rm k}}{\mathup{D} t} 
   + {\bm \nabla} \,{\bm \cdot}\, ( p {\bm v} + {\bm J}_{\rm k} ) 
   = - \rho_b \frac{\mathup{D}\Phi}{\mathup{D} t} - \rho_{0} \varepsilon_{\rm k} ,
   \label{ke_equation} 
\end{equation}
with $\rho_{0} {\bm v}\,{\bm \cdot}\,\nu\nabla^2{\bm v} = - {\bm \nabla} \,{\bm \cdot}\, {\bm J}_{\rm k} - \rho_{0}\varepsilon_{\rm k}$ 
for mechanical flux ${\bm J}_{\rm k} = -\rho_0 \nu {\bm \nabla} E_k$ 
and viscous dissipation rate $\varepsilon_{\rm k} 
= \nu (\left\lvert {\bm \nabla} u \right\rvert^2 + \left\lvert {\bm \nabla} v \right\rvert^2 + \left\lvert {\bm \nabla} w \right\rvert^2)$.
For the gravitational potential energy one obtains
\begin{equation}
   \rho_{0} \frac{\mathup{D} E_{\uphi}}{\mathup{D} t} 
   = \rho_b \frac{\mathup{D}\Phi}{\mathup{D} t} + \Phi \frac{\mathup{D}\rho_b}{\mathup{D} t} .
   \label{gpe_equation}
\end{equation}
Adding (\ref{ke_equation}) and (\ref{gpe_equation}) yields the mechanical energy budget,
\begin{equation}
   \rho_{0} \frac{\mathup{D}}{\mathup{D} t}(E_{\rm k}+E_{\uphi}) 
   + {\bm \nabla} \,{\bm \cdot}\, ( p {\bm v} + {\bm J}_{\rm k} )
   = \Phi \frac{\mathup{D}\rho_b}{\mathup{D} t} - \rho_{0}\varepsilon_{\rm k} .
   \label{mechanical_energy_budget}
\end{equation}

\subsection{Relation with compressible energetics} 
In (\ref{mechanical_energy_budget}) the right-hand side contains two
non-conservative terms: the viscous dissipation $-\rho_{0}\varepsilon_{\rm k}$ and the conversion $\Phi \mathup{D}\rho_b/\mathup{D} t$. The first is uncontroversially a transfer to internal energy, but the status of the second is less obvious. Insight follows by comparison with the compressible Navier--Stokes equations (NSE), where the conversion between kinetic and internal energies is mediated by the \emph{compressible work of expansion/contraction},
\begin{equation}
   \rho p \frac{\mathup{D}\upsilon}{\mathup{D} t} = p {\bm \nabla} \,{\bm \cdot}\, {\bm v},
   \label{compressible_work} 
\end{equation}
with $\upsilon = 1/\rho$ the specific volume. In the Boussinesq limit ${\bm \nabla}\,{\bm \cdot}\,{\bm v}=0$, so the right-hand side vanishes, yet the left-hand side remains finite when approximated with a reference pressure $p \approx -\rho_{0}\Phi$ and linearised volume
\begin{equation}
   \upsilon = \frac{1}{\rho} \approx 
   \frac{1}{\rho_{0}} - \frac{\delta \rho}{\rho_{0}^2}.
\end{equation}
This yields
\begin{equation}
   \rho p \frac{\mathup{D}\upsilon}{\mathup{D} t}
   \approx - \Phi \frac{\mathup{D} \delta \rho}{\mathup{D} t}
   = - \Phi \frac{\mathup{D}\rho_b}{\mathup{D} t} ,
\end{equation}
the second term on the right-hand side of (\ref{mechanical_energy_budget}) \citep[see][]{youngDynamicEnthalpyConservative2010}. Hence both terms indeed represent conversions with internal energy.

Although this establishes the existence of internal energy exchanges in the Boussinesq system, it does not specify the explicit form of the internal energy itself, nor does it fully validate the interpretation of $E_{\uphi}$ as gravitational potential energy. Clarifying these two points is one of the key aims of this paper.

\section{Remarks on the structure of the Navier--Stokes equations and
their energetics}
\label{sec:structure_nse} 

To understand how to construct energetically and thermodynamically consistent soundproof approximations, one must first identify where the information about energy resides in the compressible Navier--Stokes equations (NSE), and how that information may be transformed or partially suppressed when approximations are made. Following the principles recently articulated by \citet{tailleuxSimpleTransparentMethod2024}, we adopt the viewpoint that the \emph{static energy}
\[
   \Sigma(\eta,S,p,\Phi)
\]
is the fundamental scalar quantity embedding both the thermodynamics and the potential energy of the fluid. For a two-component fluid, $\Sigma$ is a function of specific entropy $\eta$, chemical composition $S$ (e.g. salinity, humidity), pressure $p$, and the geopotential $\Phi=gz$. 

\subsection{Static energy and its derivatives}

The power of this formulation lies in the fact that all relevant thermodynamic and mechanical variables appear as derivatives of $\Sigma$:
\begin{subequations}
\begin{alignat}{2}
   &\frac{\upartial \Sigma}{\upartial p} = \upsilon = \frac{1}{\rho}, 
   \qquad
   &&\frac{\upartial \Sigma}{\upartial \Phi} = B,
   \\
   &\frac{\upartial \Sigma}{\upartial \eta} = T, 
   \qquad
   &&\frac{\upartial \Sigma}{\upartial S} = \mu .
\end{alignat}
\end{subequations}
Here $\upsilon$ is the specific volume, $B$ is the quantity associated with buoyancy, $T$ is in-situ temperature, and $\mu$ the relative chemical potential. In the Boussinesq approximation $\upsilon$ is treated as constant, while in the fully compressible case it retains its variation.

In the case of compressible NSE, the static energy takes the explicit form
\begin{equation}
   \Sigma(\eta,S,p,\Phi) = h(\eta,S,p)+\Phi ,
   \label{sigma_definition}
\end{equation}
where $h(\eta,S,p)$ is the specific enthalpy. The derivatives then reduce to
\begin{subequations}
\begin{alignat}{2}
   &\frac{\upartial \Sigma}{\upartial p} = \frac{1}{\rho},
   \qquad
   &&\frac{\upartial \Sigma}{\upartial \Phi} = 1,
   \label{sigma_dynamical_derivatives}
   \\
   &\frac{\upartial \Sigma}{\upartial \eta} = T,
   \qquad
   &&\frac{\upartial \Sigma}{\upartial S} = \mu ,
   \label{sigma_thermodynamic_derivatives}
\end{alignat}   
\end{subequations}
so that $\Sigma$ unifies enthalpic and potential contributions. Potential energy is related to static energy via the relation
\begin{equation}
    E_{\rm p} = \Sigma - p \frac{\upartial \Sigma}{\upartial p} 
    = h(\eta,S,p) - \frac{p}{\rho} + \Phi .
\end{equation}
For a compressible two-component fluid, potential energy is the sum of internal energy $h-p/\rho$ and gravitational potential energy $\Phi$, as expected. For soundproof approximations, the same relation can be used to define potential energy, but the general recipe for identifying internal and gravitational potential energy is less clear and addressed in Section \ref{sec:ie_gep_partition}. 

\subsection{Dynamics in terms of $\Sigma$}

As shown by \citet{tailleuxSimpleTransparentMethod2024}, the momentum equation of NSE may be written in terms of $\Sigma$ using \eqref{sigma_dynamical_derivatives} as follows:
\begin{equation}
\begin{split} 
  & \frac{\mathup{D}{\bm v}}{\mathup{D} t} + 2 {\bm \varOmega} \times {\bm v} 
  + \frac{1}{\rho} {\bm \nabla} p + {\bm \nabla} \Phi = {\bm F} \\ 
   \Longrightarrow \quad & 
   \frac{\mathup{D}{\bm v}}{\mathup{D} t} + 2{\bm \varOmega}\times{\bm v}
   + \frac{\upartial \Sigma}{\upartial p} {\bm \nabla} p
   + \frac{\upartial \Sigma}{\upartial \Phi}{\bm \nabla} \Phi
   = {\bm F},
\end{split} 
   \label{dynamics_1}
\end{equation}
while the continuity equation becomes
\begin{equation}
\begin{split} 
   & {\bm \nabla} \,{\bm \cdot}\, {\bm v} = \frac{\mathup{D}}{\mathup{D} t} \ln{\frac{1}{\rho}} \\
   \Longrightarrow \quad & 
   {\bm \nabla} \,{\bm \cdot}\, {\bm v}
   = \frac{\mathup{D}}{\mathup{D} t}\ln\!\left(\frac{\upartial \Sigma}{\upartial p}\right),
\end{split} 
   \label{dynamics_2}
\end{equation}
where ${\bm F}$ represents viscous forces. The condition for hydrostatic balance may be written
\begin{equation}
   \frac{\upartial \Sigma}{\upartial p}{\bm \nabla} p 
   + \frac{\upartial \Sigma}{\upartial \Phi}{\bm \nabla} \Phi=0 .
\end{equation}

\subsection{Thermodynamics in terms of $\Sigma$}

The local evolution equations for entropy and salinity read
\begin{align}
   \Sigma_p^{-1}\frac{\mathup{D}\eta}{\mathup{D} t}
   &{}= -{\bm \nabla}\,{\bm \cdot}\,{\bm J}_{\eta} 
     + \Sigma_p^{-1}\dot{\eta}_{\rm irr},
   \label{local_second_law}
   \\
   \Sigma_p^{-1}\frac{\mathup{D} S}{\mathup{D} t}
   &{}= -{\bm \nabla}\,{\bm \cdot}\,{\bm J}_S ,
   \label{salinity_conservation}
\end{align}   
where $\Sigma_p^{-1}$ denotes $(\upartial\Sigma/\upartial p)^{-1}$, ${\bm J}_\eta$ and ${\bm J}_S$ are the associated diffusive fluxes, and $\dot{\eta}_{\rm irr}$ is the irreversible entropy production.

\subsection{Energy conservation and fluxes}

For the total energy
\[
   E_{\rm t} = \frac{1}{2} \left\lvert \bm{v} \right\rvert ^2 + \Sigma - p \frac{\upartial \Sigma}{\upartial p}
\]
to be conserved, one may show that the following condition is required
\begin{equation}
   \Sigma_p^{-1}\left(
      {\bm v}\,{\bm \cdot}\,{\bm F}
      + \frac{\upartial \Sigma}{\upartial \eta}\dot{\eta}
      + \frac{\upartial \Sigma}{\upartial S}\dot{S}
   \right)
   = -{\bm \nabla}\,{\bm \cdot}\,{\bm J}_E ,
\end{equation}
where ${\bm J}_E$ represents an energy flux associated with viscous and diffusive processes \citep[e.g.][]{tailleuxIdentifyingQuantifyingNonconservative2010,tailleuxObservationalEnergeticsConstraints2015}.
Here the dynamics (dependence on $p$ and $\Phi$) and thermodynamics (dependence on $\eta$ and $S$) are clearly separated.

To clarify the form of ${\bm J}_E$, first decompose the frictional work into conservative and irreversible parts,
\begin{equation}
   \Sigma_p^{-1}{\bm v}\,{\bm \cdot}\,{\bm F}
   = -{\bm \nabla}\,{\bm \cdot}\,{\bm J}_{\rm k} - \Sigma_p^{-1}\varepsilon_{\rm k},
\end{equation}
where ${\bm J}_{\rm k}$ is the mechanical flux and $\varepsilon_{\rm k}$ the viscous dissipation rate. Next, by substituting $\dot{\eta}$ and $\dot{S}$ by the right-hand sides of (\eqref{local_second_law}) and (\ref{salinity_conservation}) respectively, and separating reversible (terms expressed as the divergence of a flux) and irreversible terms, it follows that the irreversible entropy production term must be given by
\begin{equation}
   \Sigma_p^{-1}\dot{\eta}_{\rm irr}
   = \left(\frac{\upartial \Sigma}{\upartial \eta}\right)^{-1}
     \Big( \Sigma_p^{-1}\varepsilon_{\rm k}
         - {\bm J}_{\eta}\,{\bm \cdot}\,{\bm \nabla}\frac{\upartial \Sigma}{\upartial \eta}
         - {\bm J}_S\,{\bm \cdot}\,{\bm \nabla}\frac{\upartial \Sigma}{\upartial S}
     \Big),
   \label{entropy_production}
\end{equation}
with the associated total energy flux
\begin{equation}
   {\bm J}_E = 
   \frac{\upartial \Sigma}{\upartial \eta}{\bm J}_{\eta}
   + \frac{\upartial \Sigma}{\upartial S}{\bm J}_S
   + {\bm J}_{\rm k}.
   \label{energy_fluxes}
\end{equation}
See, for example, \citet{tailleuxIdentifyingQuantifyingNonconservative2010} for a detailed derivation of such results.

\subsection{Phenomenological closures}

Non-equilibrium thermodynamics prescribes linear relations between fluxes and gradients of thermodynamic potentials, for example
\begin{subequations}
\begin{align}
   &{\bm J}_{\eta} 
   = - \left( \frac{\upartial \Sigma}{\upartial \eta} \right)^{-1}
     \left( L_{\eta\eta}{\bm \nabla}\frac{\upartial\Sigma}{\upartial \eta}
          + L_{\eta S}{\bm \nabla}\frac{\upartial\Sigma}{\upartial S}
     \right), \label{eq:J_eta}
   \\
   &{\bm J}_S 
   = - \left( \frac{\upartial \Sigma}{\upartial \eta} \right)^{-1}
     \left( L_{S\eta}{\bm \nabla}\frac{\upartial\Sigma}{\upartial \eta}
          + L_{SS}{\bm \nabla}\frac{\upartial\Sigma}{\upartial S}
     \right), \label{eq:J_S}
\end{align}
\end{subequations}
\citep[e.g.][]{degrootNonequilibriumThermodynamics1962}. 
The unified static-energy formulation shows how the NSE encode energetics: every transfer pathway is mediated by derivatives of $\Sigma$. Approximations to the full system therefore amount to modifying or suppressing particular derivatives, with direct consequences for the conservation of energy and entropy. Clarifying precisely how this occurs is essential for constructing consistent soundproof models. In this paper, we will only consider the case of diffusive fluxes for
temperature and salinity that neglect cross-diffusive fluxes, see 
Appendix \ref{appC} for details. 

\section{Construction of two new models with simplified thermodynamics}
\label{sec:new_models}

In this section, we derive two new soundproof models that share the same simplified thermodynamics. The first model achieves soundproofing by simplifying the thermodynamics only, without any approximation to the dynamics. Its energetics therefore match those of the fully compressible Navier--Stokes equations (NSE). We refer to this model as the thermodynamically soundproof (TS) model. The second model uses the same thermodynamics as the TS model but adopts Boussinesq dynamics. Physically, the TS model provides an exact benchmark for assessing the Boussinesq approximation.

\subsection{Thermodynamically soundproof (TS) model}

The TS model is obtained from the NSE by simplifying the thermodynamics while keeping the equations of motion exact -- that is to say retaining the general structure of equations based on $\Sigma = h + \Phi$, but choosing an appropriately simplified expression for $h$. We seek a thermodynamic description constrained to yield a specific volume $\upsilon(S,\theta)$ that is linear in salinity $S$ and potential temperature $\theta$, and independent of pressure (implying infinite sound speed). Among possible choices, a simple approach is to define the specific enthalpy $h$ as follows:
\begin{equation}
   h(S,\theta,p) = c_p \theta + \upsilon(S,\theta) (p-p_{0}),
   \label{enthalpy_linear_eos}
\end{equation}
where $c_p$ is a constant specific heat capacity and $p_0$ a reference pressure. To express entropy in terms of $(S, \theta)$, the following expression may be derived:
\begin{equation}
   \eta(S, \theta) =
   c_p \ln{\frac{\theta}{\theta_{0}}} + \tilde{\eta}(S),
   \label{entropy_linear_eos}
\end{equation}
where $\theta_0$ is a reference potential temperature, and the function $\tilde{\eta}(S)$ encodes the saline contribution to entropy. This function is essential for ensuring irreversible entropy production from salt diffusion, which requires $\tilde{\eta}''(S)<0$, i.e. $\tilde{\eta}(S)$ must be at least quadratic in $S$ (see Appendix~\ref{appA}).

The static energy may be derived from $h$ as follows:
\begin{equation}
   \Sigma(S,\theta,p,\Phi) = c_p \theta + \upsilon(S,\theta)
   (p-p_{0}) + \Phi,
   \label{static_energy_TS}
\end{equation}
along with internal energy $E_{\rm i}$ and potential energy $E_{\rm p}$:
\begin{gather}
   E_{\rm i} 
   = h - p\upsilon 
   = c_p \theta - p_{0} \upsilon(S,\theta),
   \label{internal_energy_linear_eos}
   \\
   E_{\rm p} 
   = \Sigma - p \frac{\upartial \Sigma}{\upartial p}
   = E_{\rm i} + \Phi 
   = c_p \theta - p_{0} \upsilon(S,\theta) + \Phi.
\end{gather}
The above relations hold for any $\upsilon(S,\theta)$. To enforce linearity, we impose
\begin{equation}
   \upsilon(S, \theta) = \frac{\upartial h}{\upartial p}
   = \upsilon_{0} \left[ 1 + \alpha (\theta-\theta_{0})
   - \beta (S-S_{0}) \right],
   \label{linear_eos}
\end{equation}
with constant reference values $\upsilon_{0}$, $\theta_{0}$, $S_{0}$, and constant thermal expansion coefficient $\alpha$ and haline contraction coefficient $\beta$. Thus, the derivatives of $\Sigma$ are
\begin{subequations}
\begin{alignat}{2}
   &\frac{\upartial \Sigma}{\upartial p}
   = \upsilon_{0} \left[ 1 + \alpha (\theta-\theta_{0}) - \beta (S-S_{0}) \right],
   \qquad
   &&\frac{\upartial \Sigma}{\upartial \Phi} 
   = 1,
   \\
   &\frac{\upartial \Sigma}{\upartial \theta} 
   =
   c_p + \alpha \upsilon_{0} (p-p_{0}) , 
   \qquad
   &&\frac{\upartial \Sigma}{\upartial S}  
   = -\beta \upsilon_{0}
   (p-p_{0} ) .
\end{alignat}   
\end{subequations}

Equations~(\ref{entropy_linear_eos})--(\ref{linear_eos}) fully determine the thermodynamics \citep{iocInternationalThermodynamicEquation2010}. Although expressed in terms of $(\theta,S,p,\Phi)$, the potentials can be recast in terms of $(\eta,S,p,\Phi)$ using (\ref{entropy_linear_eos}). Using a hat to denote a representation in the $(\eta,S,p,\Phi)$ variables, e.g. $\widehat{\Sigma}(\eta, S, p, \Phi)$, equating differentials gives
\begin{equation}
   \frac{\upartial \widehat{\Sigma}}{\upartial \eta} \, \ud \eta
   + \frac{\upartial \widehat{\Sigma}}{\upartial S} \, \ud S
   = T \, \ud \eta + \mu \, \ud S
   = \underbrace{\frac{\upartial \widehat{\Sigma}}{\upartial \eta}
   \frac{c_p}{\theta}}_{\upartial \Sigma/\upartial \theta}
   \ud \theta
   + \underbrace{\left( \frac{\upartial \widehat{\Sigma}}{\upartial \eta}
   \frac{\ud \tilde{\eta}}{\ud S} + \frac{\upartial \widehat{\Sigma}}{\upartial S}
   \right)}_{\upartial \Sigma/\upartial S} \ud S ,
\end{equation}
yielding the following expressions for temperature $T$ and chemical potential $\mu$:
\begin{align}\label{Tmu_TS}
   T 
   & = \frac{\upartial \widehat{\Sigma}}{\upartial \eta} =
   \frac{\theta}{c_p} \frac{\upartial \Sigma}{\upartial \theta}
   = \theta \left( 1 +
   \frac{\alpha \upsilon_{0}(p-p_{0})}{c_p} \right)
   = \Pi(p)\,\theta ,
   \\
   \mu 
   & = \frac{\upartial \widehat{\Sigma}}{\upartial S}
   = \frac{\upartial \Sigma}{\upartial S} 
   - T \frac{\ud \tilde{\eta}}{\ud S}
   =  - \beta \upsilon_{0} (p-p_{0}) 
   - T \frac{\ud \tilde{\eta}}{\ud S} ,
   \label{t_from_theta}
\end{align}
where $\Pi(p) = 1 + \alpha \upsilon_{0} (p-p_{0})/c_p$ is analogous to the Exner function used in atmospheric science. Evaluating at the reference pressure $p=p_{0}$, yields $\Pi(p_0)=1$ and $T(p_0, \theta)=\theta$. For $p>p_{0}$, we have $T>\theta$, as in oceanic conditions.

Equation~\eqref{internal_energy_linear_eos}) shows that neglecting pressure dependence in $\upsilon$ makes internal energy a quasi-material function of $(S,\theta)$. Hence, in the absence of diffusion and viscosity, $E_{\rm i}$ is conserved along parcel trajectories and decouples from mechanical energy unless diabatic changes occur. Making the linearised specific volume independent of pressure decouples density from pressure fluctuations, thereby soundproofing the system purely through thermodynamics -- hence the TS nomenclature. The resulting velocity field has a non-zero divergence driven solely by diabatic sources in $\theta$ and $S$. This complicates numerics (as in pseudo-incompressible models, e.g. \citealt{kleinThermodynamicConsistencyPseudoincompressible2012}), motivating a Boussinesq approximation.

By substituting the appropriate derivatives of $\Sigma$ into equations \eqref{dynamics_1}, \eqref{dynamics_2}, the equations of motion for the TS model become
\begin{subequations}
\begin{gather}
   \frac{\mathup{D}{\bm v}}{\mathup{D} t} + 2 \boldsymbol{\varOmega} \times {\bm v}
   + \upsilon {\bm \nabla} p + {\bm \nabla} \Phi = {\bm F} ,
   \\
   {\bm \nabla} \,{\bm \cdot}\, {\bm v} = \frac{\mathup{D}}{\mathup{D} t} \ln{\upsilon} ,
   \\
   \rho \frac{\mathup{D}\theta}{\mathup{D} t} = - {\bm \nabla} \,{\bm \cdot}\, {\bm J}_{\theta}
   + \rho \dot{\theta}_{\rm irr},
   \label{TS_theta}
   \\
   \rho \frac{\mathup{D} S}{\mathup{D} t} = - {\bm \nabla} \,{\bm \cdot}\, {\bm J}_S .
\end{gather}   
\end{subequations}
The non-conservative production term in the $\theta$-equation is
\begin{equation}
   \rho \dot{\theta}_{\rm irr} =
   \left( \frac{\upartial \Sigma}{\upartial \theta} \right)^{-1}
   \left [ \rho \varepsilon_{\rm k} - {\bm J}_{\theta} \,{\bm \cdot}\,
   {\bm \nabla} \frac{\upartial \Sigma}{\upartial \theta}
   - {\bm J}_S \,{\bm \cdot}\, {\bm \nabla} \frac{\upartial \Sigma}{\upartial S}
   \right ] .
\end{equation}
Substituting the derivatives gives
\begin{equation}
   \rho \dot{\theta}_{\rm irr} =
   \frac{\theta}{c_p T}
   \left( \rho \varepsilon_{\rm k} - {\bm J}_{\upsilon} \,{\bm \cdot}\, {\bm \nabla} p
   \right),
\end{equation}
where ${\bm J}_{\upsilon} = \upsilon_{0}
( \alpha {\bm J}_{\theta} - \beta {\bm J}_S )$ is the specific volume flux.
As shown in Appendix \ref{appC}, 
diffusive laws relaxing toward equilibrium (${\bm \nabla} T = {\bm \nabla} \mu = {\bm 0}$) are given by
\begin{align}
   &{\bm J}_T = - \rho \kappa_T {\bm \nabla} T,
   \\
   &{\bm J}_S = - \rho \kappa_S \left( {\bm \nabla} S - \Gamma_{S,p} {\bm \nabla} p \right),
\end{align}
with $\Gamma_{S,p} = - \beta \upsilon_{0}/(T \tilde{\eta}''(S))$, neglecting cross-diffusion (Soret/Dufour effects; \citealt{greggEntropyGenerationOcean1984, greggOceanMixing2021}). From
(\ref{t_from_theta}), the potential temperature flux may be expressed as
\begin{equation}
   {\bm J}_{\theta} = \frac{1}{\Pi(p)} {\bm J}_T = \frac{\theta {\bm J}_T}{T}. 
\end{equation}

The equation for potential temperature \eqref{TS_theta} may then be written as follows:
\begin{equation}
   \rho \frac{\mathup{D}\theta }{\mathup{D} t}
   = - {\bm \nabla} \,{\bm \cdot}\, \left( \frac{\theta {\bm J}_T}{T} \right) +
   \frac{\theta}{c_p T}
   \left( \rho \varepsilon_{\rm k} - {\bm J}_{\upsilon} \,{\bm \cdot}\, {\bm \nabla} p \right).
   \label{pot_temp_TS}
\end{equation}
Appendix~\ref{appA} verifies that these closures yield non-negative irreversible entropy production. Diabatic effects induce divergence:
\begin{equation}
\begin{split}
   {\bm \nabla} \,{\bm \cdot}\, {\bm v} = \rho \frac{\mathup{D}\upsilon}{\mathup{D} t}
   &= \rho \upsilon_{0}
     \left( \alpha \frac{\mathup{D}\theta}{\mathup{D} t} - \beta \frac{\mathup{D} S}{\mathup{D} t} \right) \\
   &= - {\bm \nabla} \,{\bm \cdot}\, {\bm J}_{\upsilon}
     + \frac{\upsilon_{0} \alpha \theta}{c_p T}
     \left( \rho \varepsilon_{\rm k} - {\bm J}_{\upsilon} \,{\bm \cdot}\, {\bm \nabla} p \right) ,
\end{split}
\end{equation}
or equivalently,
\begin{equation}
   {\bm \nabla} \,{\bm \cdot}\, {\bm v} = - {\bm \nabla} \,{\bm \cdot}\, {\bm J}_{\upsilon}
   + \rho \upsilon_{0} \alpha \, \dot{\theta}_{\rm irr} .
\end{equation}
Volume integration gives
\begin{equation}
   \frac{\ud }{\ud t} V_{\rm ol}
   = \int_V {\bm \nabla} \,{\bm \cdot}\, {\bm v} \, \ud V
   = - \oint_{\upartial V} {\bm J}_{\upsilon} \,{\bm \cdot}\, {\bm n}
   \,\ud S + \int_{V} \upsilon_{0} \alpha \, \dot{\theta}_{\rm irr}
   \, \rho \, \ud V .
\end{equation}
In the absence of boundary fluxes, volume changes arise solely from $\dot{\theta}_{\rm irr}$.

\subsection{Boussinesq model}

The TS model soundproofs via thermodynamic simplification alone but leads to a computationally demanding elliptic pressure problem (Appendix~\ref{appB}). This difficulty can be eliminated via a Boussinesq approximation that simplifies the dynamics while retaining the same thermodynamics and ensuring energetic consistency.

To derive the Boussinesq model, we first make an approximation to the dynamics by choosing the approximated static energy $\Sigma_{\rm B}$ given by
\begin{equation}
   \Sigma_{\rm B}(S, \theta, p, \Phi)
   = h(S, \theta, p_{\rm{R}}(\Phi))
   + \frac{p - p_{\rm R}(\Phi)}{\rho_0}
   + \Phi
\end{equation}
with reference pressure $p_{\rm R}(\Phi) = p_{0} - \rho_{0} \Phi$ 
\citep[see][]{tailleuxSimpleTransparentMethod2024}.
Then, we simplify the thermodynamics by inserting the enthalpy as in equation \eqref{enthalpy_linear_eos}:
\begin{equation}
\begin{split}
   \Sigma_{\rm B}(S, \theta, p, \Phi) 
   &= c_p \theta + \frac{p-p_{0}}{\rho_{0}}
   + \left(2 - \rho_{0} \upsilon(S, \theta) \right) \Phi  \\
   &= c_p \theta + \upsilon_{0} \left(p-p_{0} + \rho_b (S, \theta) \Phi \right),
   \label{boussinesq_static_energy_final}
\end{split}
\end{equation}
where the density-like quantity $\rho_b$ is defined by
\begin{equation}
   \begin{aligned}
   \rho_b(S, \theta) 
   & = \rho_{0} \left( 2 - \rho_{0} \upsilon(S, \theta) \right)
   \\
   & = \rho_{0} \left( 1 - \alpha (\theta - \theta_0) +
   \beta (S - S_0)\right) .
   \end{aligned}
\end{equation}
Thus $\rho_b$ mimics classical Boussinesq density but here arises as a linear proxy for the specific volume. In the classical construction, $\rho_b$ is obtained by linearising the density equation of state; here it results from linearising the specific volume equation of state. As shown below, using $\rho_b$ simplifies the Boussinesq equations.

The potential energy associated with \eqref{boussinesq_static_energy_final} is
\begin{equation}
\begin{split}
   E_{\rm p, B} &= \Sigma_{\rm B} - p \frac{\upartial \Sigma_{\rm B}}{\upartial p} 
   \\
   &= c_p \theta - p_{0} \upsilon_{0} + \upsilon_{0}
   \rho_b\Phi .
\end{split}
\end{equation}
In general, further decomposition of potential energy into internal and gravitational components is not unique (see Section~\ref{sec:ie_gep_partition}). For the present Boussinesq model, the most natural decomposition (and the one justified in the next section) is
\begin{equation}
     E_{\rm i, B} = c_p \theta - p_{0} \upsilon_{0}, \qquad
     E_{\rm \uphi, B} = \frac{\rho_b \Phi}{\rho_{0}} ,
\end{equation}
consistent with common practice for the standard Boussinesq model.

Relevant derivatives of $\Sigma_{\rm B}$ are
\begin{subequations}
\begin{gather}
   \frac{\upartial \Sigma_{\rm B}}{\upartial p}
   = \frac{1}{\rho_{0}},
   \qquad
   \frac{\upartial \Sigma_{\rm B}}{\upartial \Phi}
   = \frac{\rho_b}{\rho_{0}},
   \\
   T_{\rm B} = \frac{\theta}{c_p} \frac{\upartial \Sigma_{\rm B}}{\upartial \theta}
   = \theta \left( 1 - \frac{\alpha \Phi}{c_p} \right) ,
   \qquad
   \mu_{\rm B} = \frac{\upartial \Sigma_{\rm B}}{\upartial S}
   - T_{\rm B} \frac{\ud\tilde{\eta}}{\ud S} = \beta \Phi - T_{\rm B} \frac{\ud \tilde{\eta}}{\ud S},
\end{gather}   
\end{subequations}
with the definition of temperature and chemical potential analogous to that of the TS model in equation \eqref{Tmu_TS}. 
Substituting the appropriate derivatives of $\Sigma_{\rm B}$ into equations \eqref{dynamics_1}, \eqref{dynamics_2}, the Boussinesq equations then read
\begin{subequations}
\begin{gather}
   \frac{\mathup{D}{\bm v}}{\mathup{D} t} + 2 \boldsymbol{\varOmega} \times {\bm v}
   + \upsilon_{0} \left( {\bm \nabla} p + \rho_b {\bm \nabla} \Phi \right) =
   \nu \nabla^2 {\bm v} ,
   \label{new_boussinesq_momentum}
   \\
   {\bm \nabla} \,{\bm \cdot}\, {\bm v} = 0 ,
   \label{new_boussinesq_continuity}
   \\
   \rho_{0} \frac{\mathup{D}\theta}{\mathup{D} t} = - {\bm \nabla} \,{\bm \cdot}\, {\bm J}_{\theta}
   + \rho_{0} \dot{\theta}_{\rm irr} ,
   \\
   \rho_{0} \frac{\mathup{D} S}{\mathup{D} t} = - {\bm \nabla} \,{\bm \cdot}\, {\bm J}_S .
\end{gather}   
\end{subequations}
These match the standard Boussinesq model with linear density in $\theta$ and $S$, but the thermodynamic interpretation is refined.
Diffusive laws are as follows:
\begin{align}
   &{\bm J}_T = - \rho_{0} \kappa_T {\bm \nabla} T_{\rm B}, 
   \qquad
   {\bm J}_{\theta} = \frac{\theta {\bm J}_T}{T_{\rm B}} ,
   \label{newB_heat_flux}
   \\
   &{\bm J}_S = - \rho_{0} \kappa_S \left( {\bm \nabla} S
   - \frac{\beta}{T_{\rm B} \, \tilde{\eta}''(S)} {\bm \nabla} \Phi \right) ,
   \label{newB_salt_flux}
\end{align}
from which we can define the flux
\begin{equation}
   {\bm J}_{\upsilon} = \upsilon_{0} \left( \alpha {\bm J}_{\theta}
   - \beta {\bm J}_S\right),
   \label{newB_specific_volume_flux}
\end{equation}
representing the flux of $v(\theta, S)$ defined in equation \eqref{linear_eos}.
Then, the non-conservative production term may be written as
\begin{equation}
   \dot{\theta}_{\rm irr} =
   \frac{\theta}{c_p T_{\rm B}}
   \left( \varepsilon_{\rm k} + {\bm J}_{\upsilon} \,{\bm \cdot}\, {\bm \nabla} \Phi
   \right) = \frac{\varepsilon_{\rm k} + {\bm J}_{\upsilon}\,{\bm \cdot}\, {\bm \nabla} \Phi}
   {c_p - \alpha \Phi}.
   \label{newB_theta_irr}
\end{equation}
Consequently, the evolution of the specific volume function $v(S, \theta)$ and density-like variable $\rho_b(S, \theta)$ is as follows:
\begin{align}
   \rho_{0} \frac{\mathup{D}\upsilon}{\mathup{D} t} 
   & =
   - {\bm \nabla} \,{\bm \cdot}\, {\bm J}_{\upsilon} + \alpha \dot{\theta}_{\rm irr} ,
   \\
   \frac{\mathup{D}\rho_b}{\mathup{D} t} 
   & = {\bm \nabla} \,{\bm \cdot}\, \left( \rho_{0} {\bm J}_{\upsilon} \right)
   - \rho_{0} \alpha \dot{\theta}_{\rm irr} .
\end{align}
Energy equations are
\begin{align}
   \rho_{0} \frac{\mathup{D} E_{\rm k}}{\mathup{D} t}  
   & = - {\bm \nabla} \,{\bm \cdot}\,
   (p \, {\bm v} + {\bm J}_{\rm k} ) 
   - \rho_b \frac{\mathup{D}\Phi}{\mathup{D} t}
   - \rho_{0} \varepsilon_{\rm k} ,
   \\
   \rho_{0} \frac{\mathup{D} E_{\uphi}}{\mathup{D} t}
   & = \rho_b \frac{\mathup{D}\Phi}{\mathup{D} t} + \Phi \frac{\mathup{D}\rho_b}{\mathup{D} t} 
   \nonumber
   \\
   & = \rho_b \frac{\mathup{D}\Phi}{\mathup{D} t} + {\bm \nabla} \,{\bm \cdot}\, \left( \rho_{0} \Phi
   {\bm J}_{\upsilon} \right) - \rho_{0} {\bm J}_{\upsilon} \,{\bm \cdot}\, {\bm \nabla} \Phi
   - \rho_{0} \alpha \Phi \, \dot{\theta}_{\rm irr} ,
   \\
   \rho_{0} \frac{\mathup{D} E_{\rm i}}{\mathup{D} t}
   & = - {\bm \nabla} \,{\bm \cdot}\, (c_p \, {\bm J}_{\theta} )
   + \rho_{0} c_p \, \dot{\theta}_{\rm irr} .
\end{align}
Summing yields the following evolution equation for total energy
\begin{equation}
   \rho_{0} \frac{\mathup{D} E_{\rm t}}{\mathup{D} t} =
   - {\bm \nabla} \,{\bm \cdot}\, {\bm J}_E + \rho_{0} E_{\rm irr} ,
\end{equation}
where the ${\bm J}_E = p \, {\bm v} + {\bm J}_{\rm k} +  \Phi \, {\bm J}_{\upsilon} + c_p \, {\bm J}_{\theta}$ is the total energy flux, and where the irreversible energy production $E_{\rm irr}$ may be expressed as:
\begin{equation}
   E_{\rm irr} = \rho_{0} (c_p - \alpha \Phi) \, \dot{\theta}_{\rm irr}
   - \rho_{0} {\bm J}_{\upsilon} \,{\bm \cdot}\, {\bm \nabla} \Phi - \rho_{0}
   \varepsilon_{\rm k} = 0 ,
\end{equation}
with the final equality resulting from \eqref{newB_theta_irr}, thus demonstrating the conservation of
${E_{\rm t} = E_{\rm k} + E_{\uphi} + E_{\rm i}}$.

\section{Is internal energy well defined in Boussinesq fluids?}
\label{sec:ie_gep_partition}

According to static energy asymptotics (SEA), an energetically and thermodynamically consistent Boussinesq approximation with a generic equation of state is obtained by approximating the static energy and potential energy as
\begin{gather}
   \Sigma_{\rm B} = h(\eta,S,p_{\rm R}(\Phi)) +
   \frac{p - p_{\rm R}(\Phi)}{\rho_{0}} + \Phi ,
   \\
   E_{\rm p,B} = \Sigma_{\rm B} - p \frac{\upartial \Sigma_{\rm B}}{\upartial p}
   = h(\eta,S,p_{\rm R}(\Phi)) - \frac{p_{\rm R}(\Phi)}{\rho_{0}} + \Phi,
\end{gather}
where the reference pressure $p_{\rm R}(\Phi)$ need not be linear in $\Phi$. In its present form, however, SEA does not prescribe how to further decompose $E_{\rm p,B}$ into internal energy (IE) and gravitational potential energy (GPE) components. Whether such a decomposition retains a unique physical meaning in a Boussinesq fluid is unclear; one must therefore rely on heuristics.

\subsection{Decomposition of potential energy}
We consider two heuristics for defining a possible decomposition of potential into IE and GPE: the first defines IE from a Legendre transform of $h$, and the second from retaining the compressible form of GPE.

\subsubsection{Heuristic I: IE as a Legendre trasform of enthalpy}
Define IE from specific enthalpy via a standard Legendre transform \citep{albertyLegendreTransformsChemical1994} as follows:
\begin{equation}
    \begin{split}
    E_{\rm i,B} 
    &= h(\eta,S,p_{\rm R}(\Phi)) - p_{\rm R}(\Phi) \, \frac{\upartial h}{\upartial p} (\eta, S, p_{\rm R}(\Phi))
    \\
    &= h(\eta,S,p_{\rm R}(\Phi))
    - \frac{p_{\rm R}(\Phi)}{\rho(\eta,S,p_{\rm R}(\Phi))} ,
    \end{split}
\end{equation}
and define GPE as the residual
\begin{equation}
   \begin{split}
      E_{\rm \uphi,B} 
      & = E_{\rm p, B} - E_{\rm i, B}
      \\
      & = \Phi + p_{\rm R}(\Phi)
      \left[ \upsilon(\eta,S,p_{\rm R}(\Phi)) - \upsilon_{0} \right] .
   \end{split}
\end{equation}
Here $E_{\rm i,B}$ retains its conventional form as in the compressible case, but $E_{\rm \uphi,B}$ contains a compressible-work-like term $p_{\rm R}(\Phi)\,(\upsilon - \upsilon_{0})$, more commonly associated with internal energy. If one further chooses a linear reference pressure, $p_{\rm R}(\Phi) = - \rho_{0} \Phi$, then
\begin{equation}
    E_{\rm \uphi,B} = \left[ 2 - \rho_{0}\,\upsilon(\eta,S,p_{\rm R}(\Phi)) \right] \Phi ,
\end{equation}
which resembles a GPE-like form. However, because part of $E_{\rm \uphi,B}$ originates from $p_{\rm R}(\Phi)$, the origin of $\Phi$ cannot be arbitrarily chosen: it must be selected so that $-\rho_{0}\Phi$ behaves as a pressure.

\subsubsection{Heuristic II: GPE as the geometric geopotential}
Define GPE as in the compressible case by setting $E_{\rm \uphi,B} = \Phi$, which then implies
\begin{equation}
    E_{\rm i,B} = h(\eta,S,p_{\rm R}(\Phi)) - 
    \frac{p_{\rm R}(\Phi)}{\rho_{0}} .
\end{equation}
In this case, $E_{\rm i,B}$ departs from the conventional form, while $E_{\rm \uphi,B}$ is the familiar geometric gravitational potential. 

\subsection{Insight from the TS model}
We lack an objective criterion to decide which heuristic is preferable in full generality. Approximating the static energy blurs the separation between internal and gravitational forms of potential energy. While the total potential energy $E_{\rm p,B}$ remains well defined its IE/GPE split is not canonical.

To gain insight, we analyse the TS model and its Boussinesq approximation. For the TS model -- whose dynamics are those of the fully compressible NSE -- the internal and gravitational potential energies are unambiguous:
\begin{alignat}{2}
   &E_{\rm i} = c_p \theta - p_{0} \upsilon
   \qquad && \implies \qquad
   {\rm IE} = \int_{V} \rho \, c_p \theta \, \ud V
   - p_{0} V_{\rm ol} ,
   \\
   &E_{\uphi} = \Phi
   \qquad && \implies \qquad
   {\rm GPE} = \int_{V} \rho \, \Phi \, \ud V ,
\end{alignat}
where $V_{\rm ol} = \int_{V} \ud V$ is the fluid volume.

As shown in Section~\ref{sec:new_models}, the potential energy for the Boussinesq approximation with pressure-independent linear equation of state is
\begin{equation}
    E_{\rm p,B} = c_p \theta - p_{0} \upsilon_{0} +
    \upsilon_{0} \rho_b(S, \theta) \, \Phi .
    \label{boussinesq_ts_ep}
\end{equation}
While the precise IE/GPE partition is not a priori unique, the structure of \eqref{boussinesq_ts_ep} naturally suggests
\begin{alignat}{2}
   &E_{\rm i,B} = c_p \theta - p_{0} \upsilon_{0}
   \qquad && \implies \qquad
   {\rm IE}_{\rm B} =
   \int_{V} \rho_{0} c_p \theta \, \ud V
   - p_{0} V_{\rm ol} ,
   \\
   &E_{\rm \uphi,B} = \upsilon_{0} \rho_b \, \Phi
   \qquad && \implies \qquad
   {\rm GPE}_{\rm B} = \int_V \rho_b \, \Phi \, \ud V .
\end{alignat}
This choice aligns with common practice in the standard Boussinesq model while maintaining traceability to the TS construction.

To assess this partition, consider an initially stratified, resting fluid relaxing toward thermodynamic equilibrium without kinetic energy generation and without external thermal, haline, or mechanical forcing. For the TS model, the energy balance reads
\begin{equation}
    \frac{\ud }{\ud t} \left( {\rm GPE} + {\rm IE} \right)
    + p_{0} \frac{\ud V_{\rm ol}}{\ud t} = 0 .
\end{equation}
Thus ${\rm IE} + {\rm GPE}$ is not conserved because the total volume changes in time, reflecting the work done by the external pressure \citep[e.g.,][]{huangAvailablePotentialEnergy2005}. Using the specific volume evolution
\begin{equation}
    \rho_{0} \frac{\mathup{D}\upsilon}{\mathup{D} t} =
    - {\bm \nabla} \,{\bm \cdot}\, {\bm J}_{\upsilon}
    + \alpha \, \dot{\theta}_{\rm irr} ,
\end{equation}
one verifies
\begin{equation}
    \frac{\ud V_{\rm ol}}{\ud t} =
    \upsilon_{0} \alpha \int_{V} \rho \, \dot{\theta}_{\rm irr} \, \ud V .
\end{equation}
Moreover, the GPE tendency becomes
\begin{equation}
    \frac{\ud }{\ud t} {\rm GPE} =
    - \int_{V} \rho \, c_p \frac{\mathup{D}\theta}{\mathup{D} t} \, \ud V
    = - \int_{V} \rho \, c_p \, \dot{\theta}_{\rm irr} \, \ud V .
\end{equation}
Hence, GPE changes are controlled by the non-conservative production/destruction of potential temperature;  even if small, $\dot{\theta}_{\rm irr}$ is fundamental for the centre-of-gravity change during equilibration. In this equilibration scenario, GPE and volume changes are proportional.

For the Boussinesq model, ${\bm \nabla} \,{\bm \cdot}\, {\bm v} = 0$, so $V_{\rm ol}$ is constant and no external-pressure work occurs. Consequently, ${\rm GPE}_{\rm B} + {\rm IE}_{\rm B}$ must remain constant:
\begin{equation}
    \frac{\ud }{\ud t} {\rm GPE}_{\rm B} =
    - \int_{V} \rho_{0} \, c_p \frac{\mathup{D}\theta}{\mathup{D} t} \, \ud V
    = - \int_V \rho_{0} \, c_p \, \dot{\theta}_{\rm irr} \, \ud V .
\end{equation}
This is identical in form to the TS case, confirming a posteriori the IE/GPE partition implied by \eqref{boussinesq_ts_ep}. As in standard Boussinesq practice, ${\rm GPE}_{\rm B} = \int \rho_b \Phi \, \ud V$. Importantly, the origin of $\Phi$ must be taken at the fluid surface because, in the Boussinesq model, $\Phi$ partly proxies the Boussinesq pressure $p_{\rm R}(\Phi) - p_{0}$ \citep[see][]{youngDynamicEnthalpyConservative2010}.
Unlike true compressible GPE, ${\rm GPE}_{\rm B}$ is not invariant under arbitrary shifts of the geopotential origin.

\section{Conclusions}
\label{sec:conclusions} 

In this paper we have revisited the construction of the Boussinesq approximation with the dual aim of restoring energetic and thermodynamic consistency, and of clarifying the physical meaning of its prognostic and diagnostic variables. The resulting model retains the familiar dynamical structure of the standard Boussinesq system, yet differs in several essential respects. In particular, density is no longer a direct thermodynamic variable but acts as a proxy for the specific volume, in-situ and potential temperature are explicitly distinguished, the diffusive salt flux is constructed to homogenise the relative chemical potential (thus correctly embedding the vertical pressure gradient), and thermodynamic potentials are made explicit rather than implicit. These modifications ensure that the model reproduces the correct irreversible behaviour for heat and salt diffusion, and that its energetics and thermodynamics remain transparently connected to those of the fully compressible Navier--Stokes equations.

To anchor the approximation, we have also introduced an exact yet practicable reference system, the TS model, which shares the same thermodynamic potentials and diffusive closures but admits a weakly divergent velocity field. This divergence, purely diabatic in origin, grows with turbulence intensity and plays a central role in the conversion pathways of turbulent mixing. Analysing the Boussinesq and TS models side by side therefore highlights the essential link between diabatic divergence and the energetics of stratified turbulence, and offers a more physically complete alternative to existing pseudo-incompressible formulations.

The analysis shows that apparently small terms -- such as those associated with the salinity-dependent part of entropy -- are indispensable carriers of fundamental information. Retaining them ensures the proper irreversibility of diffusion processes, maintains the concavity requirements of entropy, and preserves information that may be exploited by rigorous mathematical methods even if negligible in direct numerical integration. Conversely, neglecting such terms alters the structure of the equations in ways that may compromise both their energy budget and their thermodynamic admissibility.

The proposed formulation also clarifies long-standing ambiguities surrounding potential and gravitational energy in the Boussinesq approximation by confirming that it is indeed possible to interpret it as such. 

In summary, the present work provides (i) an energetically and thermodynamically consistent Boussinesq model, (ii) an exact thermodynamically soundproof model that exposes the underlying role of diabatic divergence in stratified mixing, and (iii) a framework that reconciles the simplified and non-simplified models through explicit thermodynamic potentials. While the new formulation may not radically change numerical results, it holds significant promise for theoretical analysis, offering better-defined energy budgets, clearer constraints, and new tools for understanding stratified turbulence. Future work will explore the extent to which these refinements impact rigorous bounds, mixing efficiencies, and the dynamics of geophysical and astrophysical flows.

\appendix

\section{Consistency with second law of thermodynamics} \label{appA}

Here, we verify that the assume diffusion laws are consistent with the second law of thermodynamics. In the TS model, the evolution equation for entropy may be written by taking the material derivative of the expression \eqref{entropy_linear_eos}, yielding
\begin{align}
   \rho \frac{\mathup{D}\eta}{\mathup{D} t} 
   & = 
   \frac{c_p}{\theta} \frac{\mathup{D}\theta}{\mathup{D} t}  
   + \tilde{\eta}'(S) \frac{\mathup{D} S}{\mathup{D} t}
   \\
   &= 
    \frac{c_p}{\theta} 
    \left ( - \frac{\theta}{T} {\bm \nabla} \,{\bm \cdot}\, {\bm J}_T 
    + \frac{\theta \rho \varepsilon_{\rm k}}{c_p T} 
    + \frac{\theta \beta \upsilon_{0}}{c_p T} {\bm J}_S \,{\bm \cdot}\, 
    {\bm \nabla} p \right ) - \tilde{\eta}'(S) {\bm \nabla} \,{\bm \cdot}\, {\bm J}_S ,
\end{align}
where the second equality results from a substitution of equation \eqref{pot_temp_TS}. This may be written in the form 
\begin{equation}
    \rho \frac{\mathup{D}\eta}{\mathup{D} t} = -{\bm \nabla} \,{\bm \cdot}\, {\bm J}_{\eta} 
    + \rho \dot{\eta}_{\rm irr} 
\end{equation}
where
\begin{gather}
    {\bm J}_{\eta} = \frac{c_p}{T} {\bm J}_T 
    + \tilde{\eta}'(S) {\bm J}_S ,
   \\
   \dot{\eta}_{\rm irr} = 
   \frac{c_p \kappa_T \lvert{\bm \nabla} T\rvert^2}{T^2} 
   + \frac{\varepsilon_{\rm k}}{T} 
   - \tilde{\eta}''(S) \kappa_S \lvert{\bm \nabla} S - \Gamma_{S,p} {\bm \nabla} p \rvert^2 .
\end{gather}
This expression shows that the term $\tilde{\eta}(S)$ is critical for salt diffusion to contribute to the increase in entropy, and so is the pressure dependent term via the coefficient $\Gamma_{S,p}$. It also requires that $\tilde{\eta}''(S) < 0$. 

\section{Standard diffusive laws and entropy production} \label{appC}
In this paper, we only consider diffusive laws ignoring cross-diffusive
effects (the Soret and Dufour effects, e.g., \citealt{greggEntropyGenerationOcean1984}). Here, we provide some of the
necessary background to understand how these simplified laws can be obtained.
The first step is to understand how the different fluxes for different 
variables are linked to each other, which derive 
from the total differentials for specific enthalpy and entropy
\begin{gather}
    \ud h = T \ud\eta + \mu \ud S + \upsilon \,\ud p,
    \\
    \ud\eta = \frac{c_p}{T}\, \ud T - 
    \left . \frac{\upartial \mu}{\upartial T}
    \right |_{S,p} \, \ud S - \frac{\alpha}{\rho} \, \ud p, 
\end{gather}
with the second relation providing the basis for reformulating the differential of enthalpy in terms of the $(S,T,p)$ variables:
\begin{equation}
    \ud h = c_p \ud T + 
    \left ( \mu - T \left . \frac{\upartial \mu}{\upartial T} \right |_{S,p} 
    \right ) \ud S + \upsilon (1-\alpha T) \ud p .
\end{equation}
As a result, the expressions linking the different fluxes are the same as the expressions linking the different differentials:
\begin{equation}
    {\bm J}_h = c_p {\bm J}_T + 
    \left ( \mu - T \left . \frac{\upartial \mu}{\upartial T} \right |_{S,p} \right ) 
   {\bm J}_S  = T {\bm J}_{\eta} + \mu {\bm J}_S,
\end{equation}
where
\begin{equation}
    {\bm J}_{\eta} = \frac{c_p {\bm J}_T}{T} - 
    \left . \frac{\upartial \mu}{\upartial T} \right |_{S,p} {\bm J}_S.
\end{equation}
One way to go about understanding the derivation of the phenomenological 
laws is to take as starting point the relationship
\begin{equation}
    \rho \left ( \frac{\mathup{D}h}{\mathup{D}t} - \frac{1}{\rho} \frac{\mathup{D}p}{\mathup{D}t} \right ) 
    = \rho \left ( T \frac{\mathup{D}\eta}{\mathup{D}t} + \mu \frac{\mathup{D}S}{\mathup{D}t} \right ) 
    = - {\bm \nabla} \,{\bm \cdot}\, {\bm J}_h + \rho \varepsilon_{\rm k}
\end{equation}
\citep[e.g.][]{mcdougallCommentTailleuxNeutrality2017,tailleuxObservationalEnergeticsConstraints2015,iocInternationalThermodynamicEquation2010}.
By using the above relations, assuming salinity governed by
$\rho \mathup{D}S/\mathup{D}t = - {\bm \nabla} \cdot {\bm J}_S$, it is easily established that
\begin{equation}
    \rho \frac{\mathup{D}\eta}{\mathup{D}t} = - {\bm \nabla} \,{\bm \cdot}\, {\bm J}_{\eta} 
    - \frac{c_p {\bm J}_T \,{\bm \cdot}\, {\bm \nabla} T}{T^2} 
    - \frac{1}{T} {\bm J}_S \,{\bm \cdot}\, 
    \left ( {\bm \nabla} \mu - \frac{\upartial \mu}{\upartial T} {\bm \nabla} T \right ) 
    + \frac{\rho \varepsilon_k}{T} ,
\end{equation}
which may also be written as
\begin{equation}
    \rho \frac{\mathup{D}\eta}{\mathup{D}t} = -{\bm \nabla} \,{\bm \cdot}\, {\bm J}_{\eta} 
    - \frac{c_p {\bm J}_T \,{\bm \cdot}\, {\bm \nabla} T}{T^2} 
    - \frac{1}{T} \left . \frac{\upartial \mu}{\upartial S} \right |_{T,p} 
    {\bm J}_S \,{\bm \cdot}\, \left ( {\bm \nabla} S - \Gamma_{S,p} {\bm \nabla} p \right ) 
    + \frac{\rho \varepsilon_k}{T} ,
    \label{entropy_formula} 
\end{equation}
where $\Gamma_{S,p} = - (\upartial \mu/\upartial S)^{-1} \upartial \mu/\upartial p$,
with derivatives assuming $(S,T,p)$ representation of $\mu$.
Equation (\ref{entropy_formula}) forms the basis for defining the fluxes
for ${\bm J}_T$ and ${\bm J}_S$ as follows:
\begin{gather}
    {\bm J}_T = - \rho \kappa_T {\bm \nabla} T , 
    \\
    {\bm J}_S = - \rho \kappa_S 
    \left ( {\bm \nabla} S - \Gamma_{S,p} {\bm \nabla} p\right ),
\end{gather}
which can be verified to make entropy production non-negative provided
that $\upartial \mu/\upartial S>0$.

\section{Elliptic problem for the TS model} 
\label{appB} 

Here, we briefly discuss the form of the elliptic problem satisfied by pressure in the TS model. First, owing to the decoupling between density and pressure, the continuity equation may be shown to simplify as follows
\begin{equation}
   {\bm \nabla} \,{\bm \cdot}\, ( \rho {\bm v} ) =
   - \frac{\upartial \rho}{\upartial t} 
   = \rho^2 \upsilon_{0} 
   \left( \alpha \frac{\upartial \theta}{\upartial t} 
   - \beta \frac{\upartial S}{\upartial t} \right) .
   \label{div_rhov} 
\end{equation}
Even though the right-hand side involves the Eulerian temporal derivatives of $S$ and $\theta$, \ref{div_rhov} represents a diagnostic equation for $\rho {\bm v}$, owing to 
\begin{equation}
    \frac{\upartial S}{\upartial t} = \frac{\mathup{D} S}{\mathup{D} t} - {\bm v} \,{\bm \cdot}\, 
    {\bm \nabla} S, \qquad \frac{\upartial \theta}{\upartial t} 
    = \frac{\mathup{D}\theta}{\mathup{D} t} - {\bm v} \,{\bm \cdot}\, {\bm \nabla} \theta , 
\end{equation}
being known at the present time. Next, taking the divergence of the momentum balance equation written under the form
\begin{equation}
   \frac{\upartial (\rho {\bm v})}{\upartial t} 
   + {\bm \nabla} p  = {\bm R} ,
\end{equation}
with ${\bm R} = - (\rho {\bm v} \,{\bm \cdot}\, {\bm \nabla} ) {\bm v} 
- 2 {\bm \varOmega}\times \rho {\bm v} - \rho {\bm \nabla} \Phi + \rho {\bm F}$,
is easily verified to yield the following elliptic problem for 
pressure
\begin{equation}
   \nabla^2 p 
   = - \frac{\upartial}{\upartial t} 
     \left[ \rho^2 \upsilon_{0} 
     \left( \alpha \frac{\upartial \theta}{\upartial t} 
     - \beta \frac{\upartial S}{\upartial t} \right) \right] + 
     {\bm \nabla} \,{\bm \cdot}\, {\bm R} .
   \label{elliptic_problem_for_pressure}
\end{equation}
While the estimation of ${\bm R}$ is straightforward, the time-dependent term poses challenges similar to those pertaining to pseudo-incompressible systems.

It turns out that the TS model coincides with the pseudo-incompressible model \citep{kleinThermodynamicConsistencyPseudoincompressible2012} provided the latter also assumes a pressure-independent linear equation of state. 
This may be seen by inserting the expression \eqref{enthalpy_linear_eos} for enthalpy into the pseudo-incompressible static energy ${\Sigma_{\rm PI} = h(S, \theta, p_{\rm R}(\Phi)) + v(S, \theta) (p - p_{\rm R}(\Phi)) + \Phi}$ of \citet{tailleuxSimpleTransparentMethod2024}, and seeing that it reduces to same expression as in \eqref{static_energy_TS}. Thus, the questions of numerical solutions to the TS model are of the same nature as those discussed for the pseudo-incompressible approximation.



\label{lastpage}

\end{document}